\documentclass[
  journal=pasa,
  manuscript=article-type,
  year=2025,
]{cup-journal}

\usepackage{amsmath}
\usepackage[nopatch]{microtype}
\usepackage{booktabs}
\usepackage{graphicx}
\usepackage{amssymb}
\usepackage{subcaption}
\usepackage[colorlinks=true, linkcolor=blue, citecolor=blue, urlcolor=blue]{hyperref}
\usepackage{cleveref}

\newcommand{\gr}{$\gamma$-ray}
\newcommand{\kpc}{$\mathrm{kpc}$}
\newcommand{\au}{$\mathrm{au}$}
\newcommand{\mev}{$\mathrm{MeV}$}
\newcommand{\gev}{$\mathrm{GeV}$}

\title[Microlensing of the Galactic Centre $\gamma$-Ray Excess]{Gravitational Microlensing of the Galactic
Centre $\gamma$-Ray Excess: \\ 
A New Test for Point-Like or Extended Emission?}

\author{Nada Salama}
\affiliation{Sydney Institute for Astronomy, School of Physics, The University of Sydney, 2006, NSW, Australia}
\email[N. Salama]{nada.salama@sydney.edu.au}

\author{Florian List}
\affiliation{Department of Astrophysics, University of Vienna, T\"{u}rkenschanzstra{\ss}e 17, 1180, Vienna, Austria}

\author{Geraint F. Lewis}
\affiliation{Sydney Institute for Astronomy, School of Physics, The University of Sydney, 2006, NSW, Australia}


\begin{document}

\begin{abstract}
We present a potential test of the origin of the {\gr} Galactic Centre Excess (GCE).
We demonstrate how gravitational microlensing by stellar mass objects along the line of sight to the Galactic Bulge can distinguish between the possibility of extensive emission due to dark matter self-annihilation from more prosaic astrophysical sources, namely millisecond pulsars.
Such an astrophysical origin would result in emission from a population of small, currently unresolved point-like sources -- in contrast to the expected smoother emission resulting from dark matter annihilation.
Given that the scale of gravitational microlensing, that is, the Einstein radius for stellar mass lenses, and hence, the degree of induced magnification, is sensitive 
to the size of the emitting region, such microlensing will induce time variability in the emission of astrophysical sources, whereas {\gr} emission from dark matter annihilation will effectively be immune to such influences.
However, we find that detecting microlensing-induced variability requires significantly greater sensitivity than that of current or planned {\gr} detectors.
For a small population of bright GCE sources, more than an order-of-magnitude increase in effective area over Fermi-LAT would be required, with events remaining extremely rare. For a large population of faint sources, events would occur multiple times a year, but would only be detectable with a four-order-of-magnitude improvement.
Whilst microlensing might not be a definitive test of the origin of the GCE, in future observations, it may prove useful in determining the properties of any point-like source population.

\end{abstract}

\section{Introduction}
\label{introduction}
The true nature of dark matter remains a major outstanding problem in modern physics. With the hope that dark matter interacts at least weakly with standard model particles beyond gravity, extensive ground-based experiments have been designed and performed, but have yet to yield a definitive dark matter candidate \citep{misiaszek_direct_2024}. 
Significant attention has now been turned towards the sky, with a search for the signature of dark matter physics on galactic and cosmic scales.
In particular, a pronounced enhancement of {\gr}s emitted from the Galactic Centre -- known as the GCE -- has been claimed as a potential signature of dark matter self-annihilation \citep{goodenough_possible_2009, hooper_origin_2011, hooper_dark_2011}. 

Unfortunately, the Galactic Centre is also home to complex fully astrophysical phenomena that may also be responsible for this excess, particularly a population of unresolved millisecond pulsars \citep{abazajian_consistency_2011, abazajian_astrophysical_2014, petrovic_millisecond_2015, yuan_testing_2015, oleary_young_2015, ploeg_comparing_2020}.
This has led to extensive arguments over whether the GCE is consistent with a smooth emission profile, which would be expected from dark matter annihilation or point-like astrophysical sources.

To distinguish between these two scenarios, previous works have focused on (1) the spatial morphology \citep[e.g.][]{mcdermott_morphology_2023,song_robust_2024,Muru2025}, (2) the spectrum at low \citep{hooper_pulsars_2013} and high \citep{linden_high-energy_2016} energies, and (3) photon-count statistics using
statistical \citep{lee_evidence_2016,zechlin_unveiling_2016, calore_dissecting_2021, PhysRevD.109.123042}, wavelet-based \citep{bartels_strong_2016,
zhong_new_2019}, and machine learning \citep{caron_analyzing_2018, list_galactic_2020, list_extracting_2021, list2025energy, mishra-sharma_neural_2021, caron_mind_2023, 2025PhRvD.111d3033M} methods. Recently, \cite{baxter_strategy_2023} showed that the periodicity of arriving photon counts expected from millisecond pulsars could potentially be used as an additional distinguishing feature.

In this work, we propose an entirely new approach for identifying the origin of the GCE: the emission will be subject to differing degrees of temporal magnification due to the presence of stellar mass gravitational microlenses known to lie along the line of sight to the Galactic Centre \citep{paczynski_gravitational_1991, wambsganss_discovering_1997, lewis_gravitational_2004}.
Therefore, this paper aims to demonstrate how gravitational microlensing can distinguish between extended and point-like {\gr} emission and hence between a dark matter origin or more prosaic astrophysical sources.

The structure of this paper is as follows: In Section~\ref{sec:background}, we present the background of the GCE and gravitational microlensing. In Section~\ref{sec:microlensinggce} we discuss the expected time-dependent influence of gravitational microlensing on {\gr}s from the Galactic Centre. In Section~\ref{sec:simulation}, we simulate the photon arrivals and find the plausibility of observing such temporal signatures. We present our conclusions in Section~\ref{sec:discussion}.

\section{Background}
\label{sec:background}

\subsection{Galactic Centre {\gr} Excess}
\label{subsec:gce} 

The Fermi Gamma-Ray Space Telescope has performed an all-sky survey since 2009, measuring {\gr}s with energies between $20\,$\mev$-300\,${\gev} \citep{atwood_large_2009}. 
Its primary instrument, the Large Area Telescope (LAT), has revealed an excess of {\gr}s within $10^\circ$ of the Galactic Centre, which departs from the diffuse {\gr} background and peaks at $\sim1-3$ GeV \citep{ajello_fermi-lat_2016}.

The origin of the GCE has long been a subject of debate. \cite{hooper_dark_2011} suggested that the excess is consistent with annihilating dark matter with cross section $\langle\sigma v\rangle\sim 10^{-26}\,\mathrm{cm}^3/\mathrm{s}$ and mass $m_\chi\sim7-10\,$\gev. Since then, many other works have suggested a dark matter origin \citep{hooper_origin_2011, gordon_dark_2013, abazajian_astrophysical_2014, daylan_characterization_2016, linden_high-energy_2016}. 
This is supported by the fact that the GCE is spectrally uniform across the bulge with a spectrum compatible with predicted self-annihilation spectra \citep{calore_background_2015}.
Spatially, the morphology of the excess can be fit to a modified Navarro--Frenk--White 
profile \citep[see,][]{navarro_universal_1997}, given by, 
\begin{equation}\label{eqn:NFWprofile}
    \rho(r) = \frac{\rho_0}{\left(\frac{r}{r_s}\right)^\gamma\left[1+\frac{r}{r_s}\right]^{3-\gamma}}.
\end{equation}
Here, $r$ is the radius from the Galactic Centre, $r_s$ is the scale radius, and $\rho_0$ is a characteristic density. More specifically, for dark matter pair annihilation, the resulting emission would be described by the line-of-sight integral of the squared profile from the sources to the Fermi satellite. The $\gamma$ parameter describes the slope of the inner profile. Past studies have shown that $\gamma$ ranges from $1.1$ to $1.4$ \citep{leane_spurious_2020}. 

Shortly after the GCE's discovery, a faint population of astrophysical sources was proposed as an alternative explanation. For instance, \citet{abazajian_consistency_2011} found that the GCE is spatially and spectrally consistent with known globular clusters in the Galactic Centre. These clusters are thought to host large populations of unresolved millisecond pulsars (MSPs). 
Although earlier papers found a discrepancy with MSP models in the spectrum at sub-{\gev} energies, more recent work suggests that a ``typical'' population of (potentially as many as $10^4 - 10^5$) MSPs can well explain the GCE \citep{ploeg_comparing_2020}.
Regarding the spatial morphology of the GCE, it remains debated whether a (stellar) bulge may provide a better fit than an NFW profile \citep{bartels_fermi-lat_2018,clark_dark_2018,mcdermott_morphology_2023,song_robust_2024}, and the preference has been shown to sensitively depend on the choice of employed templates.

In 2015, two independent papers suggested that the distribution is spatially clustered,
indicating the presence of astrophysical sources, 
using Non-Poissonian template fitting \citep[NPTF;][]{lee_evidence_2016, Mishra-Sharma2017} and a wavelet-based analysis \citep{bartels_strong_2016}. 
In later works using the NPTF, however, these results were called into question. \citet{leane_dark_2019} demonstrated that a smooth GCE component \textit{artificially} injected into the Fermi data was mistakenly ascribed to the point-like GCE component, indicating mismodelling (.\ \citealt{buschmann_foreground_2020, chang_characterizing_2020}). In \citet{leane_enigmatic_2020, leane_spurious_2020}, it was further shown that giving the GCE templates the freedom to float independently in the northern and southern hemispheres entirely removes the preference for point-like emission.
As for the wavelet technique, \citet{zhong_new_2019} found that when using an updated mask of resolved point sources from the 4FGL catalogue, the preference for bright GCE sources disappears (see also \citealt{zhong_robustness_2024} for a recent assessment of the robustness against masking).

More recently, convolutional neural networks have been used to disentangle the smooth and point-like components of the GCE \citep{list_galactic_2020, list_extracting_2021,mishra-sharma_neural_2021}. Although mismodelling remains a concern with these methods, the ``patch-based'' way in which they analyse photon-count maps (contrasting with the product-over-pixels likelihood employed by traditional statistical methods) results in a different response to mismodelling, offering an independent cross-check on previous analyses. 
The GCE source population preferred by these analyses would need to be significantly dimmer and consist of many more sources than the few hundred estimated by earlier works such as \citet{lee_evidence_2016}. When spectral information is additionally exploited alongside the spatial photon-count map, the GCE appears to be essentially consistent with Poisson emission \citep{list2025energy}.

In summary, 15 years after its detection, the question as to whether the GCE originates from an extended source like dark matter, or point sources such as MSPs, is still unresolved. Most analyses of the GCE to date have involved \textit{spatial} template fitting, which is particularly sensitive to the background distribution template used \citep{calore_background_2015}. 
Investigating the excess \textit{temporally} may offer an unbiased way to determine the nature of the source.

\subsection{Gravitational Microlensing}
\label{subsec:microlensing}
\citet{Paczynski_bulge_1986} suggested that gravitational microlensing could be used to probe the nature of dark matter with the Galactic Halo.
If dark matter is in the form of stellar mass compact objects, such as black holes, their gravitational influence will magnify more distant sources, resulting in a brightening and fading as the lens passes near the line of sight. 
However, even if all dark matter is in the form of stellar mass compact objects, the optical depth for significant magnification events is small ($\tau \lesssim 10^{-4}$), and it was proposed that a significant number of stellar sources, namely the Galactic Bulge and Magellanic Clouds, be targeted.
Unfortunately, after substantial effort, microlensing surveys failed to identify the expected signature of compact dark matter comprising the majority of dark matter, leading to the current focus on a particle nature. 
However, the optical depth through the Galactic Halo proved to be non-zero due to the presence of other astrophysical masses, comprising mainly of main-sequence stars, with an optical depth towards the Galactic Centre of order $\tau \sim 10^{-6}$. 

To understand the impact of gravitational microlensing, it is important to consider the relevant length scale, namely the Einstein radius in the source plane.
This is given by
\begin{equation}
R_E = \sqrt{ \frac{4 G M}{ c^2} \frac{D_{os} D_{ls} }{D_{ol}}  },
\label{eqn:einsteinradius}
\end{equation}
where $M$ is the mass of the lensing object, and $D_{ij}$ are the relative distances between the observer $(o)$, lens $(l)$ and source $(s)$. 
Given the typical lensing configuration about the Galactic Bulge, $D_{os} = 8$ \kpc\ and $D_{ol} = 4$ \kpc, this scale is
\begin{equation}
R_E \approx 8.1 \sqrt{\frac{M}{M_\odot}} \mathrm{au}.
\label{eqn:erau}
\end{equation}
For typical halo velocities of $\sim 200 \, \mathrm{km/s}$ (projected into the plane of the source), the crossing time of the Einstein radius is of order $\sim 35\,\mathrm{days}$, and hence this is the average time scale of significant magnification events. 

For a typical lens of stellar mass along the line of sight to the Galactic Bulge of $(0.1-1) \ M_{\odot}$, this scale is eight orders of magnitude larger than the $\sim 10 \ \mathrm{km}$ radius of our sources -- the MSPs. Hence, in terms of microlensing, these millisecond pulsars are effectively point-like and are sensitive to extreme gravitational lensing magnification. Emission from dark matter self-annihilation, which is expected to be smooth on \au\ scales, will be immune to the influence of gravitational microlensing.

As a lens passes along the line of sight to a source, two micro-images of the source are observed. A third image is also formed, but in such a scenario, this image is often within the lens' radius and strongly demagnified, implying that its contribution to the amplification is negligible. As is the nature of microlensing, these micro-images are too spatially clustered to be resolved. 
Instead, we observe a periodic brightening of the source, which is the contribution of the source and each micro-image. For an isolated, point-like microlensing mass, the magnification $\mu$ is given by
\begin{equation}\label{eqn:magnification}
    \mu=\frac{y^2+2}{y\sqrt{y^2+4}},
\end{equation}
where $y$ is the impact parameter \citep[see.][]{schneider_gravitational_1992}. That is the distance between the source and the lens, projected on the source plane and scaled to the Einstein radius in \Cref{eqn:einsteinradius}.

The maximum magnification occurs with perfect alignment $(y=0)$, which results in the formation of an Einstein ring. For a true point source, this magnification is formally infinite; however, our magnification is limited by the non-zero radius of the source. Taking into account the conservation of surface brightness, the magnification can be approximated as the ratio between the area of the lensed image and the area of the source. For a point source with radius $r_s \ll R_E$, the Einstein ring of radius $R_E$ has thickness $r_s$, and the ratio of areas can be approximated as
\begin{equation}
\mu_{\max}
\sim \frac{2 \pi R_E r_s}{\pi r_s^2} = 2 \frac{R_E}{r_s}.
\label{eqn:maxmag}
\end{equation}
For the scenario considered in the previous paragraph, the magnification will be of order
\begin{equation}
\mu_{\max}
  \sim 2.4 \times 10^8 \sqrt{\frac{M}{M_\odot}}.
\end{equation}\
Of course, this is the maximum magnification, and typical magnifications will be significantly lower than this upper limit. However, it is clear from stellar microlensing surveys that a proportion of microlenses within the Galactic Halo exist in binary systems, and there has even been a recent claim of a triple microlensing event towards the Galactic Bulge 
\citep{han_first_2023}. In a binary system, the region for magnification unfolds and becomes spatially extensive. Furthermore, these regions are bounded by caustics in the source plane, which are locations for formally infinite magnification for a truly point-like source, but will be limited to a very high value due to the finite nature of the source.

\subsubsection{Binary Lenses}
\label{subsec:binary}

Microlensing by a binary system will have a much stronger amplification than a single lens (SL). We consider the situation in which the stellar lens is not isolated, but in a system with another star, or perhaps a planet. To describe this system, we need to understand the mapping from the lens plane to the source plane, that is, the binary lens (BL) equation. In complex notation taken from \citet{witt_minimum_1995},
\begin{equation}
    \zeta=z
    +\frac{m_1}{\bar{w}_1-\bar{z}}
    +\frac{m_2}{\bar{w}_2-\bar{z}}.
    \label{eqn:bin_lenseq}
\end{equation}
Here, the known lens positions are $w_1$ and $w_2$, and their masses $m_1$ and $m_2$. The source position is $\zeta$ and its image is at $z$. All length scales are normalised to the Einstein radius corresponding to an arbitrary mass. We will assume later that the source positions are known, and we wish to determine the total magnification of the image. 

Inverting \Cref{eqn:bin_lenseq} to determine the image positions and the magnification is non-trivial, returning a complex 5th order polynomial \citep{witt_investigation_1990}. This polynomial will return five solutions, of which two may be imaginary. Each of the real roots corresponds to one of the five, or three, observable micro-images. As with the SL case, these images are too spatially clustered to be resolved; however, the contribution from each image is observed as the total amplification. This is the sum of the absolute amplification of each of the real solutions, 
\begin{equation} \mu_{\mathrm{tot}} 
=\sum_{i=1}^n\mu_i
=\sum_{i=1}^n|\det{J}|^{-1}|_{z=z_i}
\quad\mu_i>0,
\end{equation}
and
\begin{equation}
    \det{J}=1-\frac{d\zeta}{d\bar{z}}\frac{\overline{d\zeta}}{d\bar{z}}.
\end{equation}
Here, $z_i$ for $i = 1, \ldots, n$ denotes the position of the $i^\text{th}$ micro-image (i.e.\ the $i^\text{th}$ real root), where $n$ is 3 or 5, depending on the number of real solutions. 
Clearly, in the case that $\det{J}=0$, we obtain formally infinite amplification. This solution forms closed curves called critical curves. We can then use \Cref{eqn:bin_lenseq} to trace the image of this curve back to the source plane, obtaining the caustics.

Outside of the region bounded by the caustic, the number of real solutions to our lens equation is three. As the source and lenses move in their respective planes, the source may cross the caustic curve. When the source enters the caustic region, two things happen. Firstly, all solutions to our polynomial are real, and we observe five micro-images instead of three. Depending on the source/lens configuration, the magnification within the caustic region may be significantly larger than outside of this region. Secondly, we observe a spike in the amplification as the source crosses the caustic fold and $\det{J}=0$. This formally infinite peak is limited by the non-zero radius of the source, but given the small radius of MSPs, we expect to see a very strong peak at this point.

\begin{figure*}[t]
    \centering
    \includegraphics[width=\textwidth]{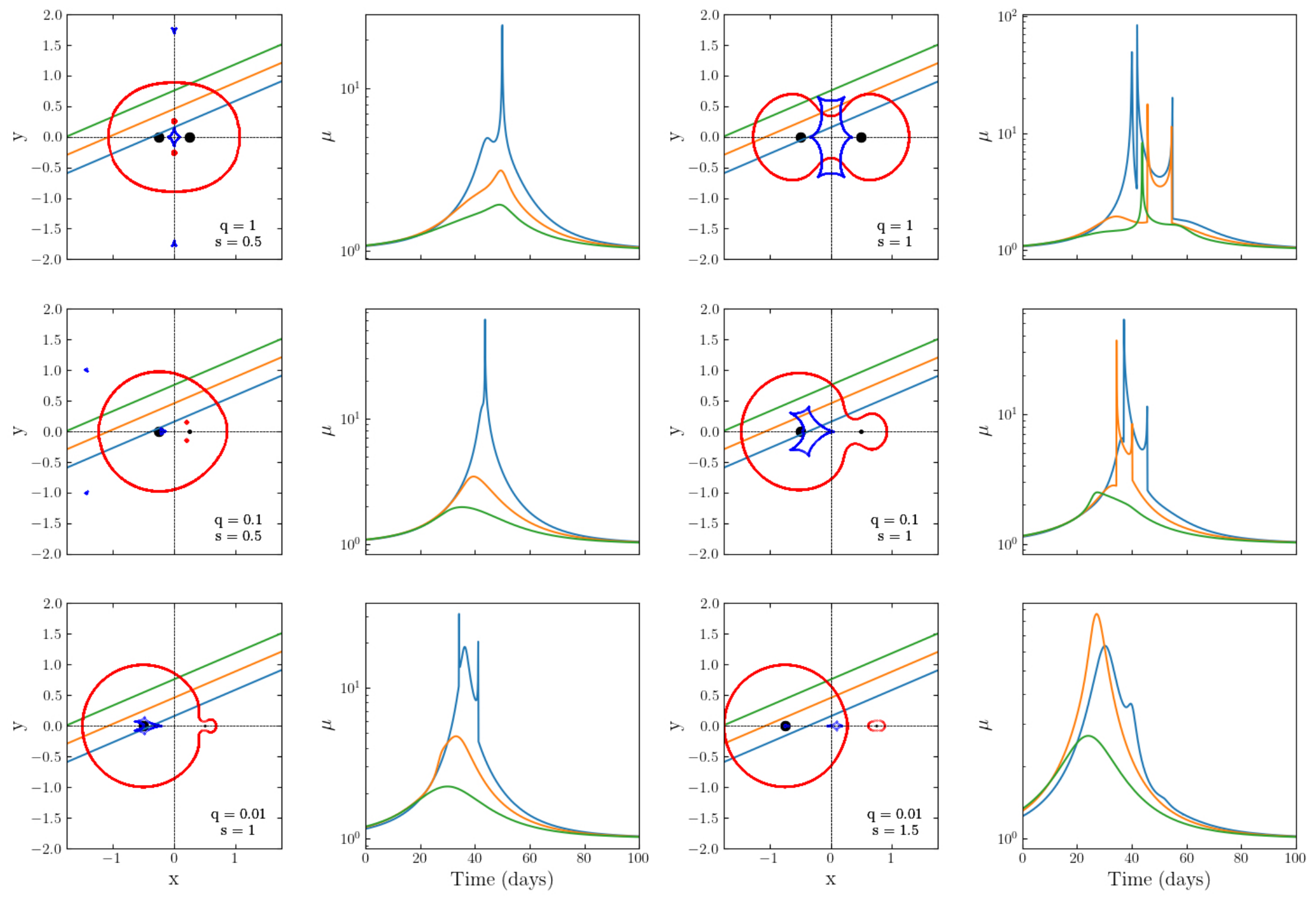}
    \caption{First and third columns (\textit{Left}) - binary lens configurations. The lenses are shown as black dots, with their relative sizes representing their mass. The source trajectories are the coloured straight lines. The critical curve is in red, and the caustic is the spiky blue curve. Second and fourth columns (\textit{Right}) - lightcurve for the respective lens configuration. The coloured lines represent the corresponding source trajectory's magnitude over time. The peaks when the source crosses a caustic are limited by the time resolution of our simulation; however, these are formally infinite.}
    \label{fig:binary}
\end{figure*}

To describe these configurations, the lenses are parametrised by two quantities, the mass ratio between the lenses ($q$), and the separation of the lenses ($s$) in units of Einstein radii. Typical examples of BL configurations are given in \Cref{fig:binary} where the lenses are the black dots, the critical curves are displayed in red, and the caustics blue. The source trajectories in the source plane are shown by the coloured lines, and the respective lightcurves are shown on their right. The three types of topologies are evident: the close BL -- which returns a central caustic and two off-axis caustics (top two panels in the left column); the intermediate BL -- with a single central caustic (bottom panel in the left column, and top two in the right column), and the wide BL -- with a central caustic and a secondary caustic (final panel). 
We will only consider a caustic crossing (CC) event in which a source enters and exits the caustic region once.

As in \Cref{eqn:maxmag}, the maximum magnification at the caustic crossings is determined by the radius of the source. Given the limiting amplitude of catalogued microlensing events, we can estimate the limiting magnification of our MSPs. Taking the maximum magnification of observed events as $\mu_\mathrm{star}$, from \citet{gaudi_gravitational_2002} the limiting magnification of our pulsars $\mu_\mathrm{MSP}$ is,
\begin{equation}\label{eqn:bin_peak}
\mu_\mathrm{MSP}\sim\sqrt{\frac{r_\mathrm{star}}{r_\mathrm{MSP}}}\mu_\mathrm{star}.
\end{equation}
Considering the microlensing events from the 2006$-$2008 seasons in the OGLE catalogue from \cite{jaroszynski_binary_2010}, we can measure the change in amplitude for the CC events. This magnitude variation is converted to a luminosity ratio, and we see an average factor of 21.6 increase in luminosity during these CC events. Of these events, some have a measured source radius (in Einstein radii). Assuming these have typical halo velocities of $200\,\mathrm{km/s}$, we can approximate that the average diameter of these sources is $\sim2\times10^9\,\mathrm{m}$. Using \Cref{eqn:bin_peak}, we obtain a limiting magnification of $\sim8700$. These events have an average impact parameter of $0.53\,R_E$; however, the impact parameter can be as large as $2\,R_E$. 

Earlier microlensing surveys identified BL and CC events by visual inspection. \citet{alcock_binary_2000} found 6.0\% of microlensing events from the MACHO survey to be caused by BLs, 66\% of these being caustic crossing. In the same year, findings from the OGLE catalogue found 9.3\% of events were attributed to BLs, 70\% being CC events. In subsequent years, the binary fraction has been reported between 1.5\% to 5.1\% \citep{jaroszynski_binary_2002, jaroszynski_binary_2004, jaroszynski_binary_2006, skowron_binary_2007, jaroszynski_binary_2010}.
It is typical to assume 6\% of microlensing events are attributed to binary lenses \citep{sumi_microlensing_2006, sumi_microlensing_2013}.
More recently, \citet{abrams_assessing_2025} used population fit models to identify BL events in lightcurves that were previously well fit by SL models, finding instead that 23.3\% of events are attributed to BLs. However, we will assume that 6\% of our lenses are BLs, and that $\sim70\%$ of these events are our CC events (so 3.5 \% of all events are CC). 

While detailed simulations are required to quantify their impact fully, we can predict that the inclusion of a third lens, due to the more intricate caustic structures, may increase the number of caustic crossings in one microlensing event. This effect is especially pronounced in non-hierarchical configurations, where all three lenses have comparable separations, leading to more complex and potentially overlapping caustics that could enhance detectability in rare cases \citep{kuang_light_2021}; these events are rare and we will not consider them further in this work.

If millisecond pulsars are the sources of the {\gr} excess at the Galactic Centre, their small size means that they would be susceptible to high magnification due to the presence of galactic halo microlensing masses. However, given the low optical depth, the frequency of any microlensing signal will depend critically on the surface density of MSPs. We turn to this in the next section.

\section{Microlensing of the GCE}
\label{sec:microlensinggce}
\begin{figure*}[htp!]
    \centering
    \includegraphics[width=\textwidth]{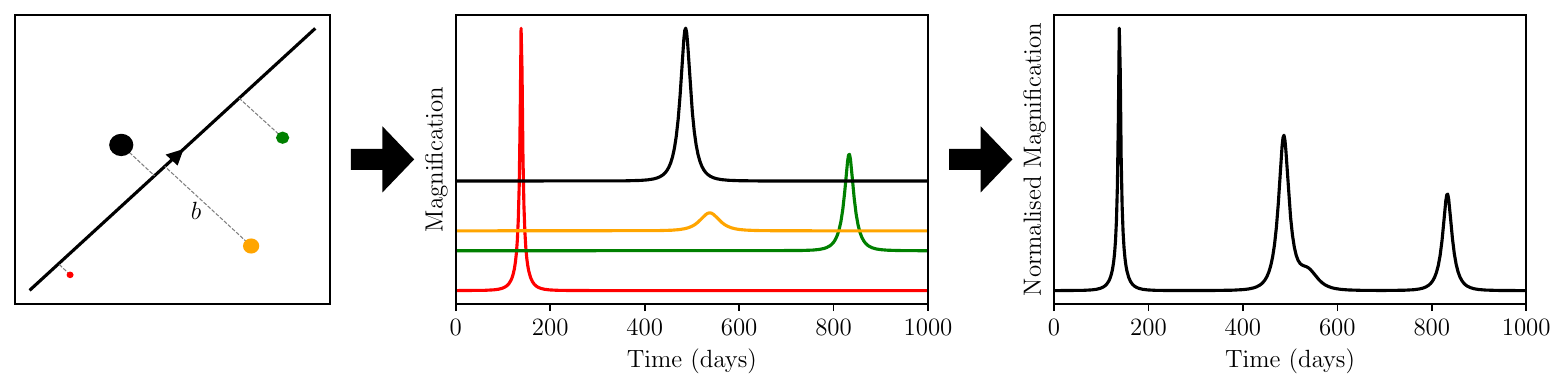}
    \caption{An illustration of the simulation process for a single lens. \textit{Left} - We generated a few sources (coloured dots), their magnitude represented by the size of the dots. A lens moves across our line of sight, its trajectory shown by the black line. \textit{Centre} - We compute the distance between each source and lens every $dt$, and calculate the magnification over time, using \Cref{eqn:magnification}. \textit{Right} - We sum all the source lightcurves and convert the result to a total magnitude representing the full sky. To quantify the amplification relative to the background, we normalise this value by the number of sources within a specified region. 
    }
    \label{fig:flowchart}
\end{figure*}
In the following, we will assume that the GCE in {\gr}s is due to a population of $N_s$ identical, point-like sources, such as MSPs. We will also assume that these sources are responsible for $N_\gamma$ photons over an observation period of $\Delta T$ years. 
A more careful treatment would model flux variations among individual MSPs via a source-count distribution. However, given that this distribution remains poorly constrained for the GCE \citep[e.g.][]{murgia_fermilat_2020}, our simple assumption of a homogeneous population is sufficient for this study. With this, the receipt of {\gr}s over the period will be given by a Poisson rate per source,
\begin{equation}
\lambda_\mathrm{LAT} = \frac{N_\gamma}{N_s \Delta T}. 
\label{eqn:poisson1}
\end{equation}

\noindent For definiteness, we consider three different scenarios here:
\paragraph{\textbf{Case 1}: $N_s = 10^5$, $\lambda_\mathrm{LAT}\sim0.01\,\mathrm{photons}/\mathrm{year}/\mathrm{source}$ \\}
Recent studies have shown that a large number of faint MSPs might be required to explain the GCE, with \cite{gautam_millisecond_2022} suggesting that accretion-induced collapse of white dwarfs could yield this number of sources. Hence, to determine the Poisson rate of the sources in our first scenario, we integrate the source-count distribution as described by $\mathrm{d}N/\mathrm{d}S$ (i.e.\ the number of sources $\mathrm{d}N$ in each infinitesimal count bin $\mathrm{d}S$) determined by \citet[Figure~5]{gautam_millisecond_2022}
over counts, which yields $N_\gamma = \int S \, \mathrm{d}N / \mathrm{d}S \, \mathrm{d}S \sim 10^4$ photons in their ROI and energy range ($|b| > 2^\circ$, $r < 25^\circ$, $2$ \gev $< E < 20$ \gev), assuming a sky exposure that corresponds to $\Delta T = 11 \, \mathrm{years}$ of Fermi data. Thus, on average, each MSP produces $0.1$ detected photon counts during that period, leading to a Poisson rate of,
\begin{equation}
\lambda_\mathrm{LAT} \sim 0.01 \ \ \mathrm{photons}/\mathrm{year}/\mathrm{source}.
\label{expect}
\end{equation}

This case maximises the probability of a microlensing event taking place, but minimises the photon rate per source, and therefore 
minimises the chance of detection for each individual lensing event due to the low Poisson rate.
    
\paragraph{\textbf{Case 2:} $N_s = 500$, $\lambda_\mathrm{LAT}\sim2\,\mathrm{photons}/\mathrm{year}/\mathrm{source}$\\}
This scenario is consistent with the bright population favoured by the NPTF analysis by \citet{lee_evidence_2016}, which would imply that some of the GCE sources lie just below the Fermi detection threshold.

A lensing event is rare, but maintaining the same total intensity of the GCE as in \citet{gautam_millisecond_2022} by determining $\lambda_{\mathrm{LAT}}$ accordingly, the photon rate per source is maximised, and so is the chance of detection.
    
\paragraph{\textbf{Case 3:} $N_s=5000$, $\lambda_\mathrm{LAT}\sim0.2\, \mathrm{photons}/\mathrm{year}/\mathrm{source}$\\} 
We chose this as our intermediate case.

With the influence of gravitational microlensing magnification, the Poisson rate is enhanced, such that the total Poisson rate for the entire population is given by
\begin{equation}
\lambda_T = \lambda_s \sum_i \mu_i,
\label{eqn:poisson}
\end{equation}
where $\mu_i$ is the magnification of the $i^\mathrm{th}$ source and $\lambda_s$ is the unlensed Poisson rate of the source.
These magnifications are temporal, depending upon the relative location of the source and the lensing masses.

To understand this temporal behaviour, it's important to remember that the optical depth $\tau(\mu=1.34)$ defines the fraction of the lensing plane within one Einstein radius (Equation~\ref{eqn:einsteinradius}), which returns a magnification of 1.34.
At any instant, the probability that at least one of $N_s$ sources lies within one Einstein radius of a lens is
\begin{equation}
p = 1 - \left( 1 - \tau \right)^{N_s}. 
\label{eqn:probability}
\end{equation}
For the GCE in MSPs with our largest considered number of sources $N_s = 10^5$ and microlensing optical depth towards the bulge of $\tau(\mu=1.34) \sim 3 \times 10^{-6}$, 
this corresponds to $p=26\%$. Note that when the magnification $\mu \gg 1$, \Cref{eqn:magnification} behaves as $\mu\approx y^{-1}$, so the resulting optical depth $\tau(\mu \gg 1 ) = \mu^{-2} \tau(\mu=1.34)$. 
The instantaneous probabilities of at least one source being 
magnified by 10 or 100 are $0.3\%$ and $0.003\%$. 
It might seem that, with these statistics, the chances of significant magnification of the GCE sources might seem unlikely, even in this optimistic case. However, the lens masses will be moving at typical halo velocities for the duration of the observations, and so the area traced out by the Einstein radius, relative to the instantaneous area, is
\begin{equation}
\frac{2 R_E v \Delta T + \pi R_E^2 }{\pi R_E^2} = \frac{2 v \Delta T}{\pi R_E} + 1
\label{eqn:area1}
\end{equation}
where $v$ is the velocity projected onto the source plane. For a solar mass lens, and assuming $v = 200 \, \mathrm{km/s}$ and $\Delta T = 11 $ years, this corresponds to a factor of 74. Hence, for the duration of the survey, the portion of the area of the source plane within one Einstein radius is $\sim 74 \tau \sim 2.4\times10^{-4}$. 
More generally, the portion of the sky contained within a distance $y$ of a lens track is given by,
\begin{equation}
A(y)=\frac{\tau}{\pi}\left[2v\Delta Ty+\pi y^2\right].
\label{eqn:totalarea}
\end{equation}
And so the total number of sources to come within a distance $y$ to a source over the survey is 
\begin{equation}
N_\mathrm{events} = A(y)N_s.
\label{eqn:nevents}
\end{equation}
So, for all microlensing events with an impact parameter less than 2, we will assume a 3.5\% probability that this lens is a BL and that the source crosses the caustic. Hence, we observe two peaks with a limiting amplification of $\sim8700$. Given the uncertainty in the source distribution and the low angular resolution for Fermi-LAT, we will consider microlensing in a small patch of sky.

\section{Simulating the microlensing signatures}
\label{sec:simulation}
\subsection{Generating a signal}
\label{subsec:lightcurve}

\begin{table}[htp!]
    \centering
    \begin{tabular}{c|c|c}
        $\mathbf{Parameter}$ & $\mathbf{Value}$ & $\mathbf{Description}$ \\ 
        \hline
        $\tau$ & $3\times10^{-6}$ & $\mathrm{Optical\ depth\ for\ microlensing}$
        \\
        $N_\mathrm{lens}$ & $6.3\times10^4$ & $\mathrm{Number\ of\ stellar\ mass\ lenses}$
        \\
        $\theta_\mathrm{region}$ & $10^\circ$ & $\mathrm{Radius\ of\ simulated\ sky\ region}$
        \\
        $R_\mathrm{region}$ & $4\times10^7R_E$ & $\mathrm{Physical\ size\ of\ region}$
        \\
        $v_\mathrm{lens}$ & $0.28\,R_E\mathrm{/day}$ & $\mathrm{Velocity\ in\ Einstein\ radii\ per\ day}$
        \\
        $y_\mathrm{max}$ & $4R_E$ & $\mathrm{Max\ impact\ parameter\ considered\ detectable}$
        \\
        $\mathrm{BL\ fraction}$ & $6\%$ & $\mathrm{Fraction\ of\ events\ caused\ by\ binaries}$
        \\
        $\mathrm{CC\ fraction}$ & $70\%\mathrm{\ of\ BLs}$ & $\mathrm{Fraction\ of\ binaries\ that\ are\ CC\ events}$
        \\
        $q,s$ & $1,1$ & $\mathrm{Mass\ ratio\ and\ separation\ for\ CC\ systems}$
    \end{tabular}
    \caption{Key parameters used in the microlensing simulations.}
    \label{tab:params}
\end{table}

\begin{figure}[t]
    \centering
    \includegraphics[width=\linewidth]{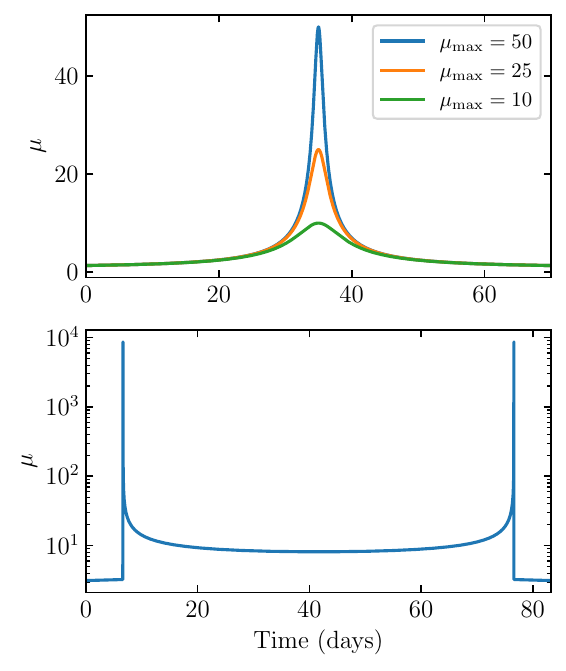}
    \caption{Examples of time-scaled lightcurves for SL \textit{(top)} and BL \textit{(bottom)} events. Depending on the impact parameter of the lensing event, we should see different maximum magnifications for the SL case.     
    }
    \label{fig:lightcurves}
\end{figure}

\begin{figure}
    \centering    
    \includegraphics[width=\linewidth]{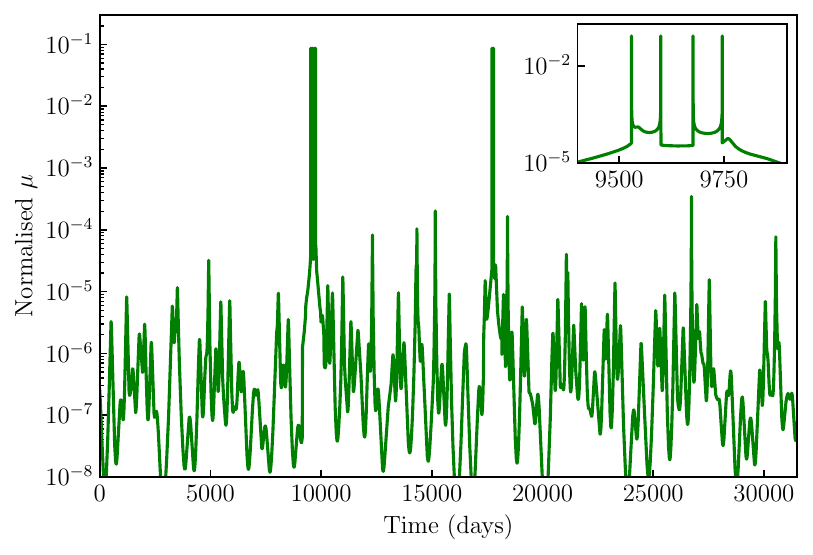}
    \caption{Simulated lightcurve of the entire Galactic Centre containing $10^5$ sources. The lightcurve is normalised to the total luminosity of $10^5$ identical sources. The inset shows the two consecutive CC events at $\sim9500$ days.}
    \label{fig:simulated_lightcurve}
\end{figure}

Our method of generating a lightcurve is summarised in Figure~\ref{fig:flowchart}, and proceeds in the following stages. 
\begin{enumerate}
    \item \textbf{Source distribution.} We simulate $N_s$ sources and randomise their distribution based on the 2D projection of the spherically symmetric NFW profile described in \Cref{eqn:NFWprofile}. A realistic MSP population will have a slightly different spatial distribution from the NFW profile \citep[e.g.][]{coleman_maximum_2020}, but for this proof of concept, this profile is sufficient. The three cases we used for $N_s$ are described in \Cref{sec:microlensinggce}.
    
    \item \textbf{Lens population.} We simulate a uniform distribution of $6.3\times10^4$ stellar-mass lenses, each assigned a 2D velocity vector with random direction and magnitude $200 \, \mathrm{km/s}$. This corresponds to $0.28\,R_E\mathrm{/day}$. The number of lenses is set by the optical depth $\tau=3\times10^{-6}$ within a lensing region with radius $10^\circ$, corresponding to $4\times10^7\,R_E$.
    
    \item \textbf{Microlensing detection.} For each source, we compute the distance to all neighbouring lenses. Based on the known magnification profile, we set a selection criterion: any lens passing within an impact parameter $y \leq 4R_E$ is considered to generate a detectable microlensing event. This threshold corresponds to a maximum peak magnification of approximately $6\%$, which we take as a conservative detection limit.

    \item \textbf{Lightcurve generation.} Each qualifying event produces a magnification lightcurve that modulates the flux of the corresponding source. Since we assume an identical brightness for all our sources, the absolute luminosity scale is arbitrary. We sum the contribution from all $N_s$ sources to produce the total observed signal over the sky.

    \item \textbf{Event classification. } In the low optical depth regime, we assume each source is lensed at most once. Of all microlensing events, we randomly designate $6\%$ as originating from BLs. Following the breakdown in \Cref{subsec:binary}, $70\%$ as classified as CC events. For simplicity, all CC events have identical lens configurations with mass ratio $q=1$ and separation $s=1$. As a result, all CC share the same lightcurve shape shown in \Cref{fig:lightcurves}. This assumption also implies uniformity in MSP radius and corresponding peak magnification. The remaining $94\%$ are classified as SL events. 
\end{enumerate}
A summary of simulation parameters can be seen in \Cref{tab:params}. An example output for $N_s=10^5$ is shown in \Cref{fig:simulated_lightcurve}. In the next section, we examine how well such a signal can be resolved by LAT, and what improvements in detector sensitivity would be needed.

\subsection{Detecting the Signatures}
To define our region of interest (ROI), we consider the distribution of sources in the Galactic Centre, based on the three cases introduced in \Cref{sec:microlensinggce}, and the angular resolution of our detector. 
The LAT has an angular resolution $\sim0.15^\circ$ at $1-3\,$\gev, assuming only the highest-graded count quartile in terms of the reconstructed direction is considered.\footnote{\url{https://www.slac.stanford.edu/exp/glast/groups/canda/lat_Performance.htm}} Hence, the
smallest region in which we can search for a microlensing event is given by a circular region with diameter $0.15^\circ$. Depending on our selected value for $N_s$, this resolution element may contain more than 1 source. We note that other event selections are also possible; for instance, lowering the requirements on the reconstructed direction would result in a larger photon number, at the expense of a worse angular resolution.

\begin{figure}[t]
    \centering
    \includegraphics[width=\linewidth]{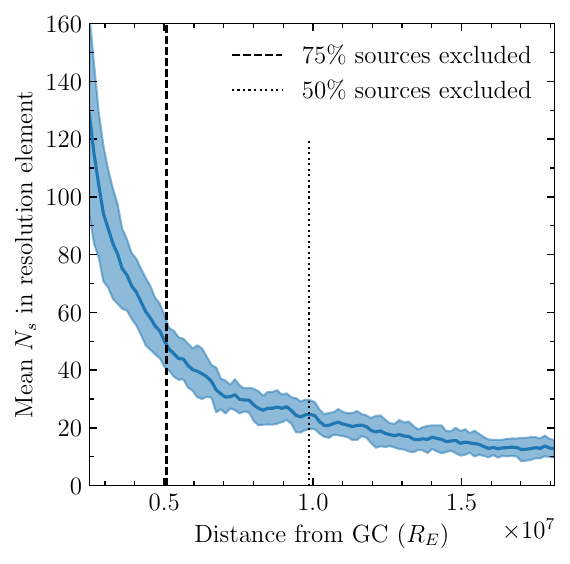}
    \caption{Average number of sources per resolution element as a function of distance from the Galactic Centre for a population of $10^5$ MSPs. The shaded region is the standard error. The dashed (dotted) line indicates the radius beyond which $75\%$ ($50\%$) of MSPs are located. 
    }
    \label{fig:n_r_element}
\end{figure}

For $N_s=10^5$, we randomly sample 15 populations of $N_s$ MSPs. 
We then count the number of sources contained within one resolution element at different distances away from the Galactic Centre, at a random angle. The average number of sources across all 15 populations within a resolution element, as a function of distance from the galactic centre, is shown in \Cref{fig:n_r_element}, with the radii containing 25\% and 50\% of the sources indicated. For the $N_s=500$ and $N_s=5000$ cases, we find that a single resolution element is extremely likely to contain only one source, so we do not apply the same treatment.

Anticipating that the LAT does not provide the sensitivity to observe these time variable signatures, we define a parameter $a$ which scales the number of photons detected by the LAT to characterise some hypothetical detector for which,
\begin{equation}
\lambda_s= a\lambda_{\mathrm{LAT}}.
\end{equation}
In our analysis, we consider only improving detector sensitivity, as next-generation {\gr} instruments are expected to improve more in this respect than in angular resolution. Improved angular resolution would reduce the number of sources within each resolution element. For a large, faint MSP population, this would lead to more isolated microlensing events per element, resembling what we will show for the $N_s=500$ and $N_s=5000$ cases. However, the individual events would still be too faint to be detected without substantially higher sensitivity. For a brighter, sparser MSP population, higher resolution would instead facilitate direct source detection.

We generate test lightcurves to determine the minimum value of $a$ that produces an observable signal. Our ROI consists of a single resolution element, populated with a realistic number of randomly distributed sources. The precise spatial distribution of sources within the ROI does not impact the statistical nature of the problem, and we assume that on these small scales, the distribution is approximately uniform anyway. We artificially modify the lightcurve of one source within the ROI to follow a microlensing profile, based on those shown in \Cref{fig:lightcurves}. The remaining sources remain unlensed, with constant magnification $\mu=1$. 

For each source, we scale the expected Poisson rate per day, $\lambda_s$, by its respective magnification curve, $\mu(t)$. We then generate the Poisson-realised lightcurve by drawing Poisson random variates for each source over the time window of interest, and combining the contributions from all sources to form the final lightcurve. The combined lightcurve is binned into 1-day-wide intervals.

To test whether an event is detectable, we compute the Poisson significance of the observed counts using the cumulative distribution function (CDF). For each 3-day sliding window, we calculate the total observed counts $k$ and the expected background $\lambda_s$ and evaluate the one-sided significance $p=1-\mathrm{CDF}(k-1;\lambda)$. This returns the probability of obtaining $k$ or more counts given the expected background. We classify an event as a significant detection if the one-sided $p$-value satisfies $p\leq0.01$.

In practice, detecting some peaks in real data would require computing the Poisson rate of each resolution element in the GC, and searching for a significant $p$-value over the entire observation period of the LAT. Of course, searching such a large parameter space significantly increases the chance of detecting high Poisson fluctuations not attributed to any microlensing events. To mitigate this so-called ``look-elsewhere effect'', one could apply a Bonferroni correction by adjusting the significance threshold based on the number of independent trials, reducing the likelihood of false positives due to multiple comparisons \citep[see][]{bayer_look-elsewhere_2020}.

While our detection method uses the Poisson CDF to evaluate the significance of observed count excesses, this implicitly assumes a fixed background rate and a fixed number of sources in each resolution element. The former assumption holds because we treat the number of contributing sources and their Poisson rates as known, except for one potential anomaly. Regarding the latter assumption, a more careful treatment would adopt a compound Poisson likelihood, which accommodates for the scatter of sources located in each resolution element; however, the effect of this correction is expected to be modest.
In practical applications, where the number of individual sources and their individual rates are not known, marginalising over the source-count distribution might be necessary.

\subsubsection{Caustic crossing event detection}

We start by considering the case in which one of our sources is lensed by a binary, and the source crosses the caustic curve. Its lightcurve is the bottom curve on \Cref{fig:lightcurves}. We should observe an amplification of $\sim8700$ at the peaks when the source crosses the caustic. These peaks only last on the order of a few minutes, and will not actually return a statistically significant increase in the number of photons arriving for small $\lambda_s$. So, we require a higher photon rate at these peaks to observe a significant signal. 

We use the method described above to generate the Poisson realised lightcurve, where one source undergoes a CC event. This is done for a range of values for $a$. For each realisation, we compute the Poisson significance in each time bin and search for two peaks separated by approximately 70 days. In practice, most lens systems along the line of sight can be assumed to have similar velocities; however, the size of the caustics depends on the specific parameters of each binary system, and thus the resulting crossing times can range from tens to hundreds of days.
A correct detection is defined as both peaks falling within 5 days of the true peaks in the microlensing lightcurve, with $p\leq0.01$. For a given $a$, we classify this correct detection as 1; otherwise, the detection value is 0. We fit a logistic curve to this data, taking the inflection point of the curve as the minimum value for $a$, which is required to detect a CC event. 

Of course, not all microlensing events will resemble the specific light curve we have simulated. Detecting such signals in real Fermi-LAT data would require placing additional constraints on the lensing configuration, including the expected peak separation and amplitude. However, such detections are not feasible with current instruments. Moreover, we do not consider the specific shape of the peaks in this analysis, as their short timescale requires a high photon count to resolve their structure. For this proof of concept, we therefore adopt an optimistic scenario to demonstrate the potential observability of the effect, and leave a more detailed treatment of the relevant parameters to future work. 

\begin{figure}[t]
    \includegraphics[width=\linewidth]{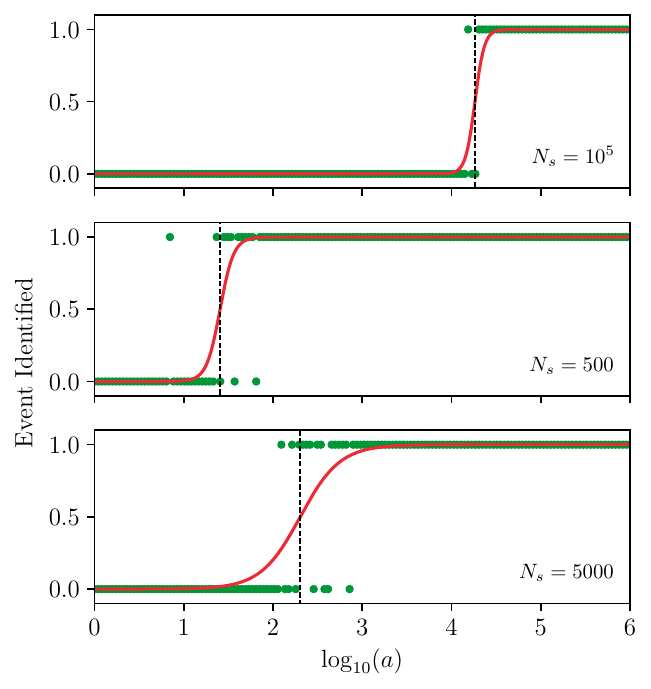}
    \caption{Correct peak detections for a CC event as a function of $\log_{10}(a)$ - the log increase in efficiency from the LAT, with the number of sources contributing to the GCE indicated on each plot. A logistic curve is fit to each of these, with the inflection defining the detector threshold; the marked inflection points from top to bottom have values $4.26$, $1.41$, $2.31$.}
    \label{fig:sigmoids}
\end{figure}

\paragraph{\textbf{Case 1}: $N_s = 10^5$, $\lambda_\mathrm{LAT}=0.01\,\mathrm{photons}/\mathrm{year}/\mathrm{source}$ \\}
Considering first the case from \cite{gautam_millisecond_2022} where $N_s$ is maximised, we show the logistic curves for detecting a CC event in our ROI containing 50 sources (\Cref{fig:sigmoids}). The required increase in detector sensitivity is a factor of $\log_{10}(a)=4.26$ to resolve the peaks on the caustic crossings. A planned future detector, such as the Advanced Particle-physics Telescope \citep{buckley_advanced_2022}, is expected to provide a magnitude increase in the effective area, and hence the detection sensitivity, with an angular resolution comparable to that of the LAT. But even then, the detection is still more than 3 orders of magnitude out of reach for the APT.

\paragraph{\textbf{Case 2:} $N_s = 500$, $\lambda_\mathrm{LAT}=2\,\mathrm{photons}/\mathrm{year}/\mathrm{source}$\\}
We now consider the opposite extreme, minimising the number of sources as in \citet{lee_evidence_2016}. Similar to the $N_s=10^5$ case, we produce artificial lightcurves and generate photon arrival times within a certain ROI. Due to the low density of sources, we can assume a ROI to contain a single source, and hence take this as the only source being microlensed. The logistic curve for detecting this event is shown in the middle panel of \Cref{fig:sigmoids}. This gives a much more optimistic value of $\log_{10}(a)=1.41$ at the inflection point, which is less than an order of magnitude larger than the sensitivity of the APT.

This method, however, does not account for the increased magnification when the source is within the caustic region. In this case, we observe an order of magnitude increase in the number of photons, lasting for $\sim 70$ days (see \Cref{fig:lightcurves}). While this small change would not be significant for the $N_s=10^5$ case, in this case, we observe 4 photons within this $\sim2$ month period, in addition to the background of $0.33$. Taking now $\log_{10}(a)=1$, we would actually see $40$ photons during this period, compared to the background rate of $4$ photons.

For a population of 500 sources, a future detector such as the APT might be able to detect the change in photon counts if a source were within the closed caustic region of a binary lens, although disentangling the microlensed GCE signal from the diffuse background might still be challenging in practice. However, the probability of observing this with the given source population is incredibly rare, the number of events being proportional to the number of sources. We predict one CC event will take place every $\sim1500$ years.

\paragraph{\textbf{Case 3:} $N_s = 5000$, $\lambda_\mathrm{LAT}=0.2\,\mathrm{photons}/\mathrm{year}/\mathrm{source}$\\}
With the statistics of the previous case being so low, let us consider an intermediate case. The counts per source are scaled down by 10 from the previous case, and we see a minimum $\log_{10}(a)=2.31$ required for the peak detection. To observe an increase from 4 to 40 photons when the source is within the caustic region, we need a $\log_{10}(a)=2$ increase in LAT's sensitivity. With this source population, one event will be expected every $\sim150$ years.

\subsubsection{Single lensing event detection}

\begin{figure*}[htp!]
    \begin{subfigure}{0.33\linewidth}
    \includegraphics[width=\linewidth]{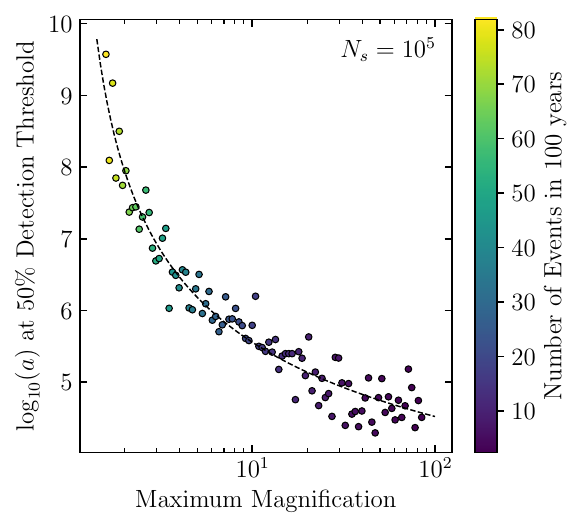}
    \end{subfigure}
    \begin{subfigure}{0.33\linewidth}
    \includegraphics[width=\linewidth]{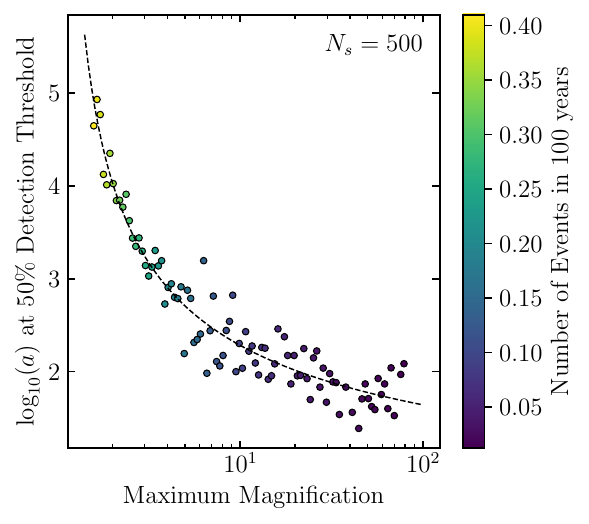}
    \end{subfigure}
    \begin{subfigure}{0.33\linewidth}
    \includegraphics[width=\linewidth]{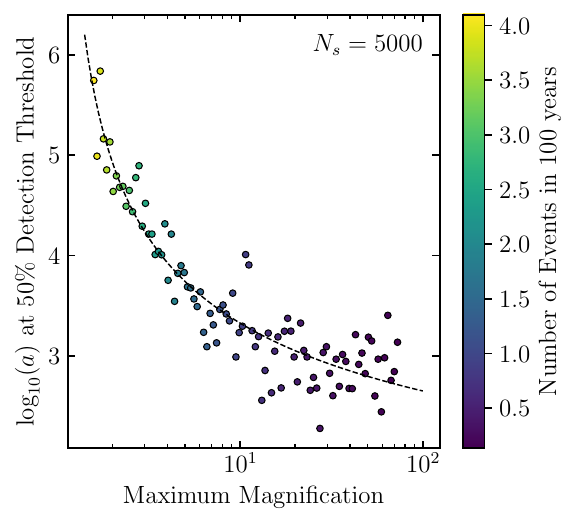}
    \end{subfigure}
    \caption{$\log_{10}(a)$ at the inflection point for the event identification curve of the SL lightcurve, as a function of maximum amplification of the SL peak. We also show a fit to these points, where the events with $\log_{10}(a)=0$ in the central panel are omitted from the fits. The colour bar indicates the predicted number of events at a given magnification over a survey of 100 years. For the left figure, the number of events is constrained to the outer 75\% of sources, where a single resolution element contains no more than 50 sources.    
    }\label{fig:SLthresholds}
\end{figure*}

We now consider the situation in which our sources are lensed by a single mass. This situation occurs more frequently and, in this case, our lightcurves peak on longer timescales, but a strong amplification is infrequent. The approximate timescale between observing strong SL events anywhere in the sky is approximately linear for $\mu\gg1$, and proportional to $N_s$ (Equation~\ref{eqn:nevents}). Here, we consider the same ROIs as in the CC case, and generate the Poisson realised lightcurve in the same way. To detect a microlensing event, we take the Poisson significance and check three criteria. First, we make sure the maximum significance is within 5 days of the closest approach between the source and lens. If this is met, we then cross-correlate the lightcurve with a family of SL templates parameterised by the impact parameter $y$. We optimise over $y$ to find the template that maximises the correlation with the observed event. If the maximum correlation exceeds 95\% and the best fit $y$ is within 20\% of the $y$ used to generate the lightcurve, we classify the event as a successful detection.

As with the previous case, this is done for a range of values of $a$, where the correct detections are classified as 1 and incorrect/no detection is 0. We fit a logistic curve to the detections as a function of $\log_{10}(a)$, and take the inflection as the minimum value of $\log_{10}(a)$ for peak detection. To determine the relationship between the strength of the peak and required $\log_{10}(a)$, we run this for a range of lightcurves, each with different $y$, and hence maximum magnification. We present the results in \Cref{fig:SLthresholds}, where the colour bar indicates the number of events of a particular magnification expected to take place in one century. 

\paragraph{\textbf{Case 1}: $N_s = 10^5$, $\lambda_\mathrm{LAT}=0.01\,\mathrm{photons}/\mathrm{year}/\mathrm{source}$ \\}

The timescale to observe a maximum amplification of 5 and 50 for $10^5$ sources is 2.5 and 25 years. This is the probability of an event taking place regardless of detection in the GCE region. If we constrain our search to the outer 75\% of sources, the timescale is 3.3 and 33.5 years, and we constrain each ROI to contain fewer than 50 sources.
The minimum $\log_{10}(a)$ for detecting a SL event, as a function of the magnification of that event, is given in \Cref{fig:n_r_element}. By minimising the number of sources within a single resolution element, we reduce the background contribution to the photon counts, improving the signal-to-noise ratio and making it easier to disentangle a potential lensing signal from background fluctuations. It is obvious here that we would still require a large enhancement to the photon rate to observe such a signal as with the CC case. 

\paragraph{\textbf{Case 2:} $N_s = 500$, $\lambda_\mathrm{LAT}=2\,\mathrm{photons}/\mathrm{year}/\mathrm{source}$\\}
Again, we take the minimum for $N_s$ and plot the value for $\log_{10}(a)$ at the inflection for the peak detection, as a function of the maximum magnification. Here we see that for stronger magnifications the required $\log_{10}(a)=2$, meaning they are not so far out of reach from the APT. 
Of course, these events are rare, with less than one per century. Considering even the weaker microlensing events are still incredibly rare, and will require a large increase in the sensitivity of our detector.

\paragraph{\textbf{Case 3:} $N_s = 5000$, $\lambda_\mathrm{LAT}=0.2\,\mathrm{photons}/\mathrm{year}/\mathrm{source}$\\}
For the intermediate case, we should observe around one SL event per century, and our required photon count is three orders of magnitude larger than Fermi-LAT.

\section{Conclusions}
\label{sec:discussion}

We have explored the susceptibility of the Galactic Centre {\gr} excess to gravitational microlensing due to the presence of stellar mass lenses along the line of sight to the Galactic Bulge. If the source of this excess is astrophysical, in particular, due to emission from millisecond pulsars, we expect these to be strongly magnified during microlensing events. However, smooth {\gr} emission, expected from dark matter annihilation, will be immune to the influence of this gravitational lensing.

We determined the frequency of these events based on observed microlensing events of known astrophysical sources within our own Galaxy. Although infrequent, we know that some of these events will be attributed to binary lenses. Given that the number of sources believed to contribute to the GCE is not well constrained, the number of microlensing events is uncertain. Detecting these microlensing events and understanding their frequency may place constraints on the number of MSPs in the galactic centre.

In order to determine the plausibility of detecting these signatures, we simulated the photon arrival times from these events. We showed that the LAT is not sensitive enough to observe most of these signatures and identified a factor $a$ increase in the efficiency of a future {\gr} detector, which allows these signals to be detected from the background. We found that an upcoming instrument like the APT, which is predicted to have $\log_{10}(a)=1$, is still insufficient to resolve CC and SL even for optimistic source populations like $N_s=500$ or $5000$. That said, given that the source-count distribution for these rather bright populations would peak only slightly below the current Fermi detection threshold, a significant fraction of MSPs would be expected to be individually resolved anyway with these more powerful detectors. For instance, \cite{Dinsmore_luminosity_2021} estimate that increasing the detection sensitivity by a factor of $10-20\times$ would lead to $20 - 30\%$ of the GCE being resolved even in their most pessimistic scenarios.

For a more extreme scenario involving $10^5$ faint sources, we estimate that a detector with $\log_{10}(a)>4.26$ might begin detecting these events, but even then, they remain rare, occurring perhaps a few times a century. Assuming the MSPs are identical, a single detected microlensing event could still offer insights into the number of sources contributing to the GCE, though such detections will likely require next-generation instruments with significantly greater sensitivity.

It is important to note that we have assumed the only time variation would be due to microlensing events. 
The Fermi all-sky variability analysis (FAVA) suggests that for a sizeable fraction of {\gr} sources, there are intrinsic flux variations on the order of weeks \citep{ackermann_fermi_2013}. Although most flaring sources appear to be extragalactic, recent findings do suggest some of them may be associated with known sources in the Galactic Centre region \citep{joffre_historical_2024}. However, the shape of these lightcurves is not explicitly characterised, and it is not clear how these signals would differ from the microlensing-induced variability at such low photon counts. As such, in the event of a potential microlensing detection, the possibility of an intrinsic flare would need to be carefully considered as an alternative hypothesis.

Ultimately, we show that a larger population of MSPs would result in more frequent microlensing events, though these would remain undetectable with current {\gr} technology. Conversely, a smaller MSP population may produce microlensing signals within the reach of next-generation {\gr} detectors. Our study demonstrates the potential of gravitational microlensing as a powerful probe of compact object populations in high-background environments, including those beyond the visible spectrum.

\paragraph{Acknowledgements}
NS acknowledges funding from the University of Sydney through the Faculty of Science Research Stipend Scholarship. The authors thank Nick Rodd for comments on an earlier draft of this work. FL thanks Christopher Eckner, Noemi Anau Montel, and Francesca Calore for insightful discussions on the GCE. 

\paragraph{Data Availability Statement}
The data generated for this paper will be made available on reasonable request to the corresponding author.

\paragraph{Author Contributions}
Conceptualisation: G.F.L. Methodology: N.S; G.F.L; F.L. Data curation: N.S. Data visualisation: N.S. Writing original draft: N.S; F.L; G.F.L. All authors approved the final submitted draft.

\printendnotes

\bibliography{references}

\end{document}